%% file: cobstat.tex
\documentstyle[prl,aps,twocolumn,psfig]{revtex}
\newskip\humongous \humongous=0pt plus 1000pt minus 1000pt

\newif\ifdtup

\relax

\newcommand{\newc}{\newcommand}

\newc{\be}{\begin{equation}}
\newc{\ee}{\end{equation}}
\newc{\ba}{\begin{eqnarray}}
\newc{\ea}{\end{eqnarray}}
\newc{\bea}{\begin{eqnarray}}
\newc{\eea}{\end{eqnarray}}
\newc{\D}{\partial}
\newc{\ie}{{\it i.e.} }
\newc{\eg}{{\it e.g.} }
\newc{\etc}{{\it etc.} }
\newc{\etal}{{\it et al.}} 

\newc{\ra}{\rightarrow}
\newc{\lra}{\leftrightarrow}
\newc{\lsim}{\buildrel{<}\over{\sim}}
\newc{\gsim}{\buildrel{>}\over{\sim}}

\begin{document}

\title{Hints of Cosmic String Induced Discontinuities in the COBE Data?}

\bigskip

\author{L. Perivolaropoulos, N. Simatos}
  
\address{Department of Physics,
University of Crete\\ 
P.O.Box 2208 ,GR-710 03 Heraklion, Crete; Greece\\
e-mail: leandros@physics.uch.gr, simatos@physics.uch.gr}

\date{\today}
\maketitle

\begin{abstract}
We apply new statistical tests on the four year 53 GHz DMR data and show that there
is significant probability for the existence of coherent temperature discontinuities
hidden in the CMB maps around the Galactic Poles. Comparing Monte Carlo
simulations with the DMR maps we find that the probability for the existence
of a coherent discontinuity with amplitude $0.2 \times ({{\delta T} \over
T})_{rms}$ superposed on the gaussian fluctuations is more than double than the
corresponding probability when no discontinuity is present. This result is
consistent with the existence of a horizon size long cosmic string with mass
per unit length $\mu$ given by $G\mu v_s \gamma_s \simeq 10^{-6}/\pi$
where $v_s$ is the string velocity and $\gamma_s$ is the corresponding
Lorentz factor.
  
\end{abstract}

\narrowtext

\bigskip

\section{INTRODUCTION}
\noindent

New statistical tests have recently been proposed for the detection of
coherent discontinuities (edges) hidden in CMB maps \cite{lean}. The main
advantage of these statistics is that they focus  on the large scale
coherence properties of CMB maps and are therefore effective even in cases of
low resolution maps provided that the sky area covered is large. These
properties are shared by the maps produced by the DMR instrument of COBE. The goal of this paper
is the application of the new statistical tests analyzed in Ref. \cite{lean}
on the four year DMR maps in the regions of the Galactic Poles
(approximatelly $90^{\circ} \times 90^{\circ}$ or $32 \times 32$ pixel maps). The
statistics calculated for these maps are the skewness, the kurtosis, the
Sample Mean Difference (SMD) \cite{lean} and the Maximum Sample Difference
(MSD) \cite{lean}. These results are then compared with the corresponding
results obtained from a large number of gaussianized DMR maps (similar power
spectrum and random phases in Fourier space) with and without a coherent
discontinuity superposed. Thus for each applied statistic we find the value of
the superposed coherent discontinuity amplitude $\frac{\delta T}{T} = \alpha
\times (\frac{\delta T}{T})_{rms}$
that is most consistent with
the actual DMR data.
 
In particular we ask the following question: `Assume that the DMR map around
the North Galactic Pole is gaussian but has small coherent temperature discontinuities
superposed on it. What is the most probable value of the amplitude of the
superposed  discontinuity? Also, what is the ratio of the probability that a
discontinuity of the given amplitude is present over the probability that no
discontinuity is present?'. Clearly, for a gaussian map we would find that the
most probable value of the discontinuity amplitude is 0.

Before describing the technique followed to address the above questions we
will briefly review the statistical tests we have used (the notation used
here is the same as in Ref.\cite{lean}). These
tests involve both conventional statistics (skewness and kurtosis) and new
statistics (MSD and SMD) optimized for the detection of coherent
discontinuities in 1d and 2d pixel maps. Assuming a $32 \times 32$
standardized temperature pixel map $T_{ij}$
($i,j=1,...,32$), the skewness $s$ and the kurtosis $k$ are defined as 

\ba
s&=&<T^3>\,\equiv \sum_{i,j} T_{ij}^3/(32)^{2} \\
k&=&<T^4>\,\equiv \sum_{i,j} T_{ij}^4/(32)^{2}
\ea
These are conventional statistics and their values for gaussian maps with
uncorrelated pixels are $s=0$, $k=3$. To define the statistics MSD and SMD we
consider a partition of the CMB in two parts separated by a random line. In
this study we have considered straight lines but the analysis can be
generalized to more general types of lines with no significant changes in the
results. Let ${\bar k}$ denote the set of parameters that define the
partition line and let ${\bar T}_u$ and ${\bar T}_l$ be the mean temperatures
of the two parts of the map (the indices `$u$' and `$l$' stand for `upper'
and `lower' parts). The statistical variable $Y_{\bar k}$ is defined as:
\ba
Y_{\bar k} \equiv {\bar T}_u-{\bar T}_l    
\ea
The statistics Sample Mean Difference (SMD) and Maximum Sample Difference
(MSD) are defined as :
\ba
SMD &=&  \frac{1}{N} \sum_{\bar k} Y_{\bar k} \\
MSD &=& \max (Y_{\bar k})                        
\ea
where $N$ is the total number of partitions. Both MSD and SMD approach
asymptotic values for large N.  

Consider now a gaussian pattern of temperature fluctuations with a small
coherent temperature discontinuity defined by a partition $\bar k_{0}$
superposed on the map with coherence scale comparable to the size of the pattern. In
Ref.\cite{lean} it was shown that the presence of the coherent discontinuity
$\bar k_{0}$ can be detected much more efficiently by the statistics SMD
and MSD than by the skewness and
kurtosis. A physically motivated mechanism which can lead to the production of
a coherent discontinuity on CMB maps is the presence of a moving long cosmic
string in our horizon \cite{shelard}. 

The main mechanism by which strings can produce CMB fluctuations on angular
scales larger than $1^{\circ}-2^{\circ}$  
has been well studied both analytically 
\cite{veer90} and
simulated numerically \cite{ben92} and is known as the {\it
Kaiser-Stebbins effect} \cite{kais84}. 
According to this effect, moving long strings present between the time of recombination
$t_{rec}$ 
and the present time $t_0$, produce step-like temperature discontinuities between photons that
reach the observer through
opposite sides of the string. These discontinuities are due to the peculiar 
nature of the spacetime
around a long string which even though is {\it locally} flat, {\it globally}
has 
the geometry of
a cone with deficit angle $8\pi G\mu$ ($G$ is Newton's constant, $\mu$ is the mass per unit 
length of the string and we have used units with $c=1$). The magnitude of the discontinuity is 
proportional to the
deficit angle, to the string velocity $v_s$ and depends on the relative 
orientation between the unit
vector along the string ${\hat s}$ and the unit photon wave-vector ${\hat
k}$. It is given by \cite{veer90} :
\begin{equation} 
{{\delta T}\over T}=\pm 4\pi G\mu v_s \gamma_s {\hat k} \cdot ({\hat
v_s}\times 
{\hat s})
\end{equation}
where $\gamma_s$ is the relativistic Lorentz factor and the sign changes when 
the string is
crossed. The angular scale over which this discontinuity persists is given by 
the radius of
curvature of the string which according to simulations 
\cite{ben88} is approximately equal
to the horizon scale. 
The growth of the horizon from $t_{rec}$ to $t_0$ results in a superposition
of a large number of step-like temperature seeds of all sizes starting from
about $2^{\circ}$ (the angular size of the horizon at $t_{rec}$) to about 
$180^{\circ}$ (the present horizon scale). By the Central Limit Theorem (CLT)
this
large number of superposed seeds results in a pattern of fluctuations that
obeys gaussian statistics. Thus the probability distribution for the 
temperature of each pixel of a CMB map with resolution larger than about
$1^{\circ} - 2^{\circ}$ is a gaussian  \cite{coul94}. 

However, on large angular scales, despite the large number of superposed seeds there 
is also coherence of induced fluctuations. The
fluctuations
on these scales may be viewed as a superposition of a gaussian scale 
invariant background coming mainly from small scale seeds plus a small 
number of step-like discontinuities which are coherent and persist on angular
scales larger than $100^{\circ}$. The
 presence of coherent discontinuities due to late long strings, superposed on a
gaussian background appears to be a generic feature of cosmic string models. In what 
follows we will explore the consequences on non-gaussianity of this generic feature of 
the cosmic string models.

\section{STATISTICAL TESTS}

Most of the statistical tests for the detection of non-gaussianity focus on small scale
properties (e.g. peaks) of CMB maps, because it is usually thought that the
CLT tends to hide non-gaussianity on large scales due to the large number of
superposed seeds on these scales. This expectation, however, is
incorrect for the type of non-gaussianity induced by moving late long
strings. The basic feature of this type of non-gaussian fluctuations is the large scale
coherence and can thus be best detected by a specially designed statistical
test. Such are the tests SMD and MSD analysed both analytically and
numerically in Ref.\cite{lean}.

Here, we apply these statistics along with the more conventional skewness and
kurtosis on the four year 53 GHz DMR data. To find the probability that the 
DMR maps (around the Galactic Poles) are gaussian with a superposed
discontinuity of a given amplitude $\frac{\delta T}{T} = \alpha \times
(\frac{\delta T}{T})_{rms}$, we apply the following technique:

\begin{enumerate}
\item We consider two $32 \times 32$ pixel maps of the 53 GHz (A+B)/2 DMR
      data (monopole and dipole removed, optimum Galactic cut: 3881 surviving
      pixels \cite{ben96}) around the Galactic
      Poles\footnote[1]{As North and South Galactic Pole regions we have used
      faces 0 and 5 of the COBE quadrilateralized spherical cube projection
      (CSC) in galactic pixelization, according to the plots and pixel
      numbering schemes found at:

      http://www.gsfc.nasa.gov/astro/cobe/skymap\_info.html.}
      (about $ 90^{\circ} \times 90^{\circ} $ on the sky). 
      We calculate the statistics: skewness, kurtosis, SMD and MSD.
\item We construct a large number (1200) of gaussianized maps obtained from
      the DMR map by randomizing the phases in Fourier space and using a
      gaussian spectrum with 0 mean and variance equal to the 
      DMR measured spectrum.
\item For each one of the constructed maps we calculate the four statistics:
      skewness, kurtosis, SMD and MSD.
\item We superpose on the constucted maps a random coherent
      discontinuity of amplitude $\alpha \times (\frac{\delta T}{T})_{rms}$,
      obtained by a straight line partition and recalculate the skewness,
      kurtosis, SMD, and MSD statistics.
\item Using the results from steps 3 and 4 we construct the probability
      distribution
      for the skewness, kurtosis, SMD and MSD for various values of $\alpha$
      (e.g. Figs. 1, 2).
\item We use the results from steps 1 and 5 to find the probability for
      obtaining the already calculated, from step 1, COBE DMR values of skewness,
kurtosis, SMD and MSD in the
      gaussianized maps as a function of $ \alpha $. Thus we have four
      probability distributions, one for each statistic: $ P_{s}(\alpha),
      P_{k}(\alpha),P_{SMD}(\alpha),P_{MSD}(\alpha) $ (Figs. 3, 4). 
\end{enumerate}

For convenience we will hereafter use normalized probability distributions
defined as:
\ba
Q(\alpha) \equiv \frac{P(\alpha)}{P(\alpha=0)} 
\ea
$ Q(\alpha) $ helps us answer the
question: `How probable is the existence of a coherent discontinuity within
the studied maps?'. 
Fig. 1 shows the probability distribution of the skewness,
for various values of $\alpha$. Clearly, this statistic is
insensitive to the detection of a steplike coherent
discontinuity in the data. The situation is similar for the kurtosis. 

\begin{figure}[ht]
\centerline{
\input{plotskew.latex}
}
\caption{
Skewness probability distribution of the 1200 gaussianized maps, obtained from
the COBE DMR data around the North Galactic Pole, for two values of the
discontinuity amplitude $\alpha$, ($\alpha$=0 continuous line, $\alpha$=0.2
dashed line). The two dots represent the probability of getting the actual
value of the skewness rendered by the DMR data for each value of $\alpha$.
}
\end{figure}
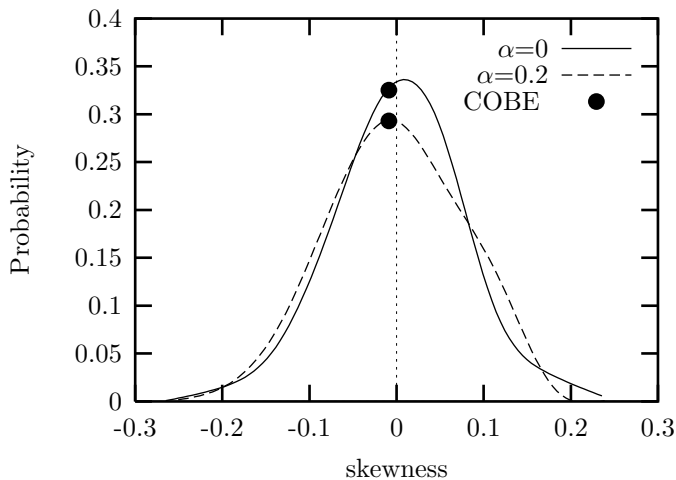

Thus, these two conventional statistics cannot differentiate between a gaussian
map and a map with a small superposed coherent discontinuity with $\alpha \leq
0.8 $.
This is clearly seen in Figs. 3, 4 where $Q_{s}(\alpha), Q_{k}(\alpha )$ are shown to be 
quite flat functions of $\alpha$.

The situation is quite different for the MSD and SMD
statistics. The probability distribution for the MSD, for example, is clearly
very sensitive to the value of $\alpha$ and its maximum occurs for $\alpha =
0.2 $, thus favouring the existence of a discontinuity (Fig. 2).

\begin{figure}[ht]
\centerline{
\input{plotmsd.latex}
}
\caption{
MSD probability distribution around the North Galactic Pole (see caption in
Fig.1).
}
\end{figure}
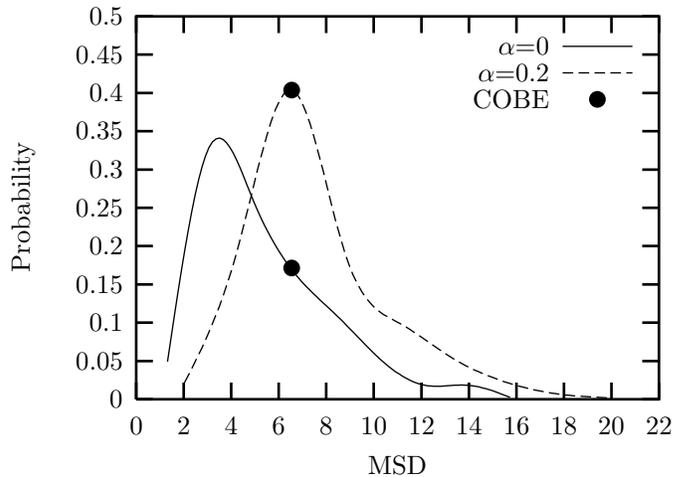

Clearly, the
distributions $Q_{MSD}(\alpha)$ and $Q_{SMD}(\alpha)$, as it is shown in
Figs. 3, 4, demonstrate a conspicuous peak at $\alpha \simeq 0.2 $  with
$ \frac{Q(\alpha=0.2)}{Q(\alpha=0)} \simeq 2.4 $ indicating that the presence
of a discontinuity with $ \alpha \simeq 0.2 $ is more than twice as
probable
compared to a purely gaussian map. We therefore conclude that there are
clear hints for the existence of a temperature discontinuity in the 53 GHz
(A+B)/2 DMR data in the region of the North Galactic Pole superposed on the
gaussian fluctuations (Fig. 3).

\begin{figure}[ht]
\centerline{
\input{plotnorth.latex}
}
\caption{Smoothed normalized probability distributions of
the COBE DMR values calculated by the four statistics as a function
of the discontinuity amplitude
$\alpha$. The distributions where obtained from the
1200 gaussianized maps covering the North Galactic Pole region.
}
\end{figure}
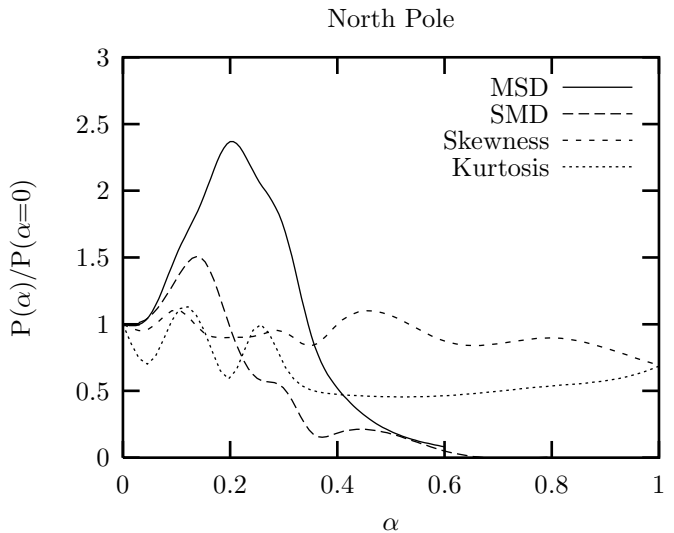

 We have repeated the analysis for the data in the
region of the South Galactic Pole (Fig. 4) with very similar results,
indicating the presence of a similar type of discontinuity.

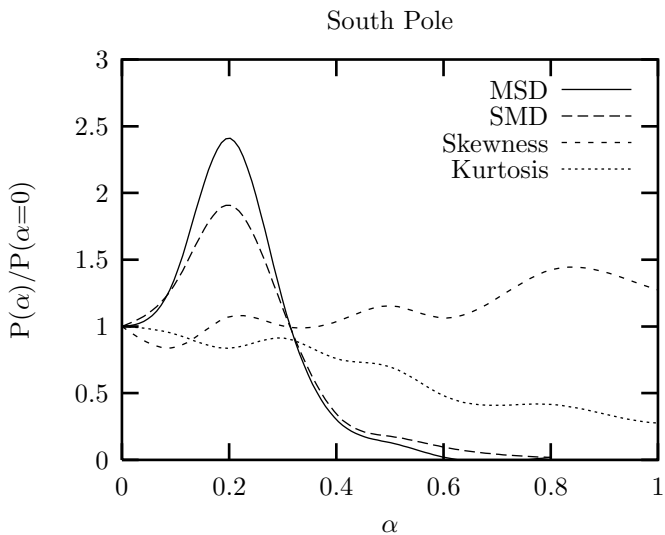
\begin{figure}[ht]
\centerline{
\input{plotsouth.latex}
}
\caption{Smoothed normalized probability distributions, South Galactic
Pole region (see caption in Fig. 3).
}
\end{figure}

The presence of such discontinuities on top of the anticipated gaussian
fluctuations could be due to the presence of a long string in our present
horizon. According to the results from the application of the MSD
statistic, such a string would go approximately through the galactic
coordinates ($230^{\circ}.98  ,71^{\circ}.68 $) and ($ 330^{\circ}.25 ,
48^{\circ}.98 $)
in the North Galactic Pole
region (or pixels 612 and 29 according to the quadrilateralized spherical
cube). The corresponding
discontinuity in the South Galactic Pole goes through the coordinates
($ 315^{\circ}.00 , -73^{\circ}.57 $) and ($ 122^{\circ}.02 , -73^{\circ}.12
$) or pixels 5733 and 5553 .
It is straightforward to check that the position of this discontinuity
relative to the dipole anisotropy excludes the possibility that the
discontinuity is an artifact of improper removal of the dipole anisotropy. 
       
\section{CONCLUSIONS}

We have shown that the 53 GHz (A+B)/2 DMR data in the regions around the
Galactic Poles show clear hints for the existence of hidden temperature
discontinuities coherent on large angular scales. These hints have been
revealed by the application of the new MSD and SMD statistics which were
especially designed for the detection of coherent discontinuities hidden in
the fluctuation patterns. We have also shown that the conventional statistics 
of skewness and kurtosis are unable to reveal these features. The 53 GHz DMR maps 
in the regions around the
Galactic Poles are consistent with the presence of no discontinuities at the
$1\sigma$ level, but the probability for the existence of a coherent
discontinuity with amplitude $\frac{\delta T}{T} = \alpha \times
(\frac{\delta T}{T})_{rms} \simeq 4 \times 10^{-6}$ is more than double 
compared to the probability that
no discontinuity is present. This result could be interpreted as an
indication for the existence of a moving long string in our present horizon
with $G\mu v_s \gamma_s \simeq 10^{-6}/\pi $.

\section{ACKNOWLEDGEMENTS}

We would like to thank Dr. G. Hinshaw for his assistance in acquiring and
interpreting the COBE DMR data.
This work is the result of a network supported by the European Science
Foundation. The European Science Foundation acts as a catalyst for the
development of science by bringing together leading scientists and funding
agencies to debate, plan and implement pan-European activities.
 This work was also supported by the EU grant CHRX-CT94-0621 as well as by the
Greek General Secretariat of Research and Technology grant пемед95-1759.

\end{document}

%% file: plotskew.latex
\setlength{\unitlength}{0.1bp}
\special{!
/gnudict 120 dict def
gnudict begin
/Color false def
/Solid false def
/gnulinewidth 5.000 def
/userlinewidth gnulinewidth def
/vshift -33 def
/dl {10 mul} def
/hpt_ 31.5 def
/vpt_ 31.5 def
/hpt hpt_ def
/vpt vpt_ def
/M {moveto} bind def
/L {lineto} bind def
/R {rmoveto} bind def
/V {rlineto} bind def
/vpt2 vpt 2 mul def
/hpt2 hpt 2 mul def
/Lshow { currentpoint stroke M
  0 vshift R show } def
/Rshow { currentpoint stroke M
  dup stringwidth pop neg vshift R show } def
/Cshow { currentpoint stroke M
  dup stringwidth pop -2 div vshift R show } def
/UP { dup vpt_ mul /vpt exch def hpt_ mul /hpt exch def
  /hpt2 hpt 2 mul def /vpt2 vpt 2 mul def } def
/DL { Color {setrgbcolor Solid {pop []} if 0 setdash }
 {pop pop pop Solid {pop []} if 0 setdash} ifelse } def
/BL { stroke gnulinewidth 2 mul setlinewidth } def
/AL { stroke gnulinewidth 2 div setlinewidth } def
/UL { gnulinewidth mul /userlinewidth exch def } def
/PL { stroke userlinewidth setlinewidth } def
/LTb { BL [] 0 0 0 DL } def
/LTa { AL [1 dl 2 dl] 0 setdash 0 0 0 setrgbcolor } def
/LT0 { PL [] 0 1 0 DL } def
/LT1 { PL [4 dl 2 dl] 0 0 1 DL } def
/LT2 { PL [2 dl 3 dl] 1 0 0 DL } def
/LT3 { PL [1 dl 1.5 dl] 1 0 1 DL } def
/LT4 { PL [5 dl 2 dl 1 dl 2 dl] 0 1 1 DL } def
/LT5 { PL [4 dl 3 dl 1 dl 3 dl] 1 1 0 DL } def
/LT6 { PL [2 dl 2 dl 2 dl 4 dl] 0 0 0 DL } def
/LT7 { PL [2 dl 2 dl 2 dl 2 dl 2 dl 4 dl] 1 0.3 0 DL } def
/LT8 { PL [2 dl 2 dl 2 dl 2 dl 2 dl 2 dl 2 dl 4 dl] 0.5 0.5 0.5 DL } def
/Pnt { stroke [] 0 setdash
   gsave 1 setlinecap M 0 0 V stroke grestore } def
/Dia { stroke [] 0 setdash 2 copy vpt add M
  hpt neg vpt neg V hpt vpt neg V
  hpt vpt V hpt neg vpt V closepath stroke
  Pnt } def
/Pls { stroke [] 0 setdash vpt sub M 0 vpt2 V
  currentpoint stroke M
  hpt neg vpt neg R hpt2 0 V stroke
  } def
/Box { stroke [] 0 setdash 2 copy exch hpt sub exch vpt add M
  0 vpt2 neg V hpt2 0 V 0 vpt2 V
  hpt2 neg 0 V closepath stroke
  Pnt } def
/Crs { stroke [] 0 setdash exch hpt sub exch vpt add M
  hpt2 vpt2 neg V currentpoint stroke M
  hpt2 neg 0 R hpt2 vpt2 V stroke } def
/TriU { stroke [] 0 setdash 2 copy vpt 1.12 mul add M
  hpt neg vpt -1.62 mul V
  hpt 2 mul 0 V
  hpt neg vpt 1.62 mul V closepath stroke
  Pnt  } def
/Star { 2 copy Pls Crs } def
/BoxF { stroke [] 0 setdash exch hpt sub exch vpt add M
  0 vpt2 neg V  hpt2 0 V  0 vpt2 V
  hpt2 neg 0 V  closepath fill } def
/TriUF { stroke [] 0 setdash vpt 1.12 mul add M
  hpt neg vpt -1.62 mul V
  hpt 2 mul 0 V
  hpt neg vpt 1.62 mul V closepath fill } def
/TriD { stroke [] 0 setdash 2 copy vpt 1.12 mul sub M
  hpt neg vpt 1.62 mul V
  hpt 2 mul 0 V
  hpt neg vpt -1.62 mul V closepath stroke
  Pnt  } def
/TriDF { stroke [] 0 setdash vpt 1.12 mul sub M
  hpt neg vpt 1.62 mul V
  hpt 2 mul 0 V
  hpt neg vpt -1.62 mul V closepath fill} def
/DiaF { stroke [] 0 setdash vpt add M
  hpt neg vpt neg V hpt vpt neg V
  hpt vpt V hpt neg vpt V closepath fill } def
/Pent { stroke [] 0 setdash 2 copy gsave
  translate 0 hpt M 4 {72 rotate 0 hpt L} repeat
  closepath stroke grestore Pnt } def
/PentF { stroke [] 0 setdash gsave
  translate 0 hpt M 4 {72 rotate 0 hpt L} repeat
  closepath fill grestore } def
/Circle { stroke [] 0 setdash 2 copy
  hpt 0 360 arc stroke Pnt } def
/CircleF { stroke [] 0 setdash hpt 0 360 arc fill } def
/C0 { BL [] 0 setdash 2 copy moveto vpt 90 450  arc } bind def
/C1 { BL [] 0 setdash 2 copy        moveto
       2 copy  vpt 0 90 arc closepath fill
               vpt 0 360 arc closepath } bind def
/C2 { BL [] 0 setdash 2 copy moveto
       2 copy  vpt 90 180 arc closepath fill
               vpt 0 360 arc closepath } bind def
/C3 { BL [] 0 setdash 2 copy moveto
       2 copy  vpt 0 180 arc closepath fill
               vpt 0 360 arc closepath } bind def
/C4 { BL [] 0 setdash 2 copy moveto
       2 copy  vpt 180 270 arc closepath fill
               vpt 0 360 arc closepath } bind def
/C5 { BL [] 0 setdash 2 copy moveto
       2 copy  vpt 0 90 arc
       2 copy moveto
       2 copy  vpt 180 270 arc closepath fill
               vpt 0 360 arc } bind def
/C6 { BL [] 0 setdash 2 copy moveto
      2 copy  vpt 90 270 arc closepath fill
              vpt 0 360 arc closepath } bind def
/C7 { BL [] 0 setdash 2 copy moveto
      2 copy  vpt 0 270 arc closepath fill
              vpt 0 360 arc closepath } bind def
/C8 { BL [] 0 setdash 2 copy moveto
      2 copy vpt 270 360 arc closepath fill
              vpt 0 360 arc closepath } bind def
/C9 { BL [] 0 setdash 2 copy moveto
      2 copy  vpt 270 450 arc closepath fill
              vpt 0 360 arc closepath } bind def
/C10 { BL [] 0 setdash 2 copy 2 copy moveto vpt 270 360 arc closepath fill
       2 copy moveto
       2 copy vpt 90 180 arc closepath fill
               vpt 0 360 arc closepath } bind def
/C11 { BL [] 0 setdash 2 copy moveto
       2 copy  vpt 0 180 arc closepath fill
       2 copy moveto
       2 copy  vpt 270 360 arc closepath fill
               vpt 0 360 arc closepath } bind def
/C12 { BL [] 0 setdash 2 copy moveto
       2 copy  vpt 180 360 arc closepath fill
               vpt 0 360 arc closepath } bind def
/C13 { BL [] 0 setdash  2 copy moveto
       2 copy  vpt 0 90 arc closepath fill
       2 copy moveto
       2 copy  vpt 180 360 arc closepath fill
               vpt 0 360 arc closepath } bind def
/C14 { BL [] 0 setdash 2 copy moveto
       2 copy  vpt 90 360 arc closepath fill
               vpt 0 360 arc } bind def
/C15 { BL [] 0 setdash 2 copy vpt 0 360 arc closepath fill
               vpt 0 360 arc closepath } bind def
/Rec   { newpath 4 2 roll moveto 1 index 0 rlineto 0 exch rlineto
       neg 0 rlineto closepath } bind def
/Square { dup Rec } bind def
/Bsquare { vpt sub exch vpt sub exch vpt2 Square } bind def
/S0 { BL [] 0 setdash 2 copy moveto 0 vpt rlineto BL Bsquare } bind def
/S1 { BL [] 0 setdash 2 copy vpt Square fill Bsquare } bind def
/S2 { BL [] 0 setdash 2 copy exch vpt sub exch vpt Square fill Bsquare } bind def
/S3 { BL [] 0 setdash 2 copy exch vpt sub exch vpt2 vpt Rec fill Bsquare } bind def
/S4 { BL [] 0 setdash 2 copy exch vpt sub exch vpt sub vpt Square fill Bsquare } bind def
/S5 { BL [] 0 setdash 2 copy 2 copy vpt Square fill
       exch vpt sub exch vpt sub vpt Square fill Bsquare } bind def
/S6 { BL [] 0 setdash 2 copy exch vpt sub exch vpt sub vpt vpt2 Rec fill Bsquare } bind def
/S7 { BL [] 0 setdash 2 copy exch vpt sub exch vpt sub vpt vpt2 Rec fill
       2 copy vpt Square fill
       Bsquare } bind def
/S8 { BL [] 0 setdash 2 copy vpt sub vpt Square fill Bsquare } bind def
/S9 { BL [] 0 setdash 2 copy vpt sub vpt vpt2 Rec fill Bsquare } bind def
/S10 { BL [] 0 setdash 2 copy vpt sub vpt Square fill 2 copy exch vpt sub exch vpt Square fill
       Bsquare } bind def
/S11 { BL [] 0 setdash 2 copy vpt sub vpt Square fill 2 copy exch vpt sub exch vpt2 vpt Rec fill
       Bsquare } bind def
/S12 { BL [] 0 setdash 2 copy exch vpt sub exch vpt sub vpt2 vpt Rec fill Bsquare } bind def
/S13 { BL [] 0 setdash 2 copy exch vpt sub exch vpt sub vpt2 vpt Rec fill
       2 copy vpt Square fill Bsquare } bind def
/S14 { BL [] 0 setdash 2 copy exch vpt sub exch vpt sub vpt2 vpt Rec fill
       2 copy exch vpt sub exch vpt Square fill Bsquare } bind def
/S15 { BL [] 0 setdash 2 copy Bsquare fill Bsquare } bind def
/D0 { gsave translate 45 rotate 0 0 S0 stroke grestore } bind def
/D1 { gsave translate 45 rotate 0 0 S1 stroke grestore } bind def
/D2 { gsave translate 45 rotate 0 0 S2 stroke grestore } bind def
/D3 { gsave translate 45 rotate 0 0 S3 stroke grestore } bind def
/D4 { gsave translate 45 rotate 0 0 S4 stroke grestore } bind def
/D5 { gsave translate 45 rotate 0 0 S5 stroke grestore } bind def
/D6 { gsave translate 45 rotate 0 0 S6 stroke grestore } bind def
/D7 { gsave translate 45 rotate 0 0 S7 stroke grestore } bind def
/D8 { gsave translate 45 rotate 0 0 S8 stroke grestore } bind def
/D9 { gsave translate 45 rotate 0 0 S9 stroke grestore } bind def
/D10 { gsave translate 45 rotate 0 0 S10 stroke grestore } bind def
/D11 { gsave translate 45 rotate 0 0 S11 stroke grestore } bind def
/D12 { gsave translate 45 rotate 0 0 S12 stroke grestore } bind def
/D13 { gsave translate 45 rotate 0 0 S13 stroke grestore } bind def
/D14 { gsave translate 45 rotate 0 0 S14 stroke grestore } bind def
/D15 { gsave translate 45 rotate 0 0 S15 stroke grestore } bind def
/DiaE { stroke [] 0 setdash vpt add M
  hpt neg vpt neg V hpt vpt neg V
  hpt vpt V hpt neg vpt V closepath stroke } def
/BoxE { stroke [] 0 setdash exch hpt sub exch vpt add M
  0 vpt2 neg V hpt2 0 V 0 vpt2 V
  hpt2 neg 0 V closepath stroke } def
/TriUE { stroke [] 0 setdash vpt 1.12 mul add M
  hpt neg vpt -1.62 mul V
  hpt 2 mul 0 V
  hpt neg vpt 1.62 mul V closepath stroke } def
/TriDE { stroke [] 0 setdash vpt 1.12 mul sub M
  hpt neg vpt 1.62 mul V
  hpt 2 mul 0 V
  hpt neg vpt -1.62 mul V closepath stroke } def
/PentE { stroke [] 0 setdash gsave
  translate 0 hpt M 4 {72 rotate 0 hpt L} repeat
  closepath stroke grestore } def
/CircE { stroke [] 0 setdash 
  hpt 0 360 arc stroke } def
/BoxFill { gsave Rec 1 setgray fill grestore } def
end
}
\begin{picture}(2520,1944)(0,0)
\special{"
gnudict begin
gsave
0 0 translate
0.100 0.100 scale
0 setgray
newpath
LTb
500 400 M
63 0 V
1907 0 R
-63 0 V
500 581 M
63 0 V
1907 0 R
-63 0 V
500 761 M
63 0 V
1907 0 R
-63 0 V
500 942 M
63 0 V
1907 0 R
-63 0 V
500 1122 M
63 0 V
1907 0 R
-63 0 V
500 1303 M
63 0 V
1907 0 R
-63 0 V
500 1483 M
63 0 V
1907 0 R
-63 0 V
500 1664 M
63 0 V
1907 0 R
-63 0 V
500 1844 M
63 0 V
1907 0 R
-63 0 V
500 400 M
0 63 V
0 1381 R
0 -63 V
828 400 M
0 63 V
0 1381 R
0 -63 V
1157 400 M
0 63 V
0 1381 R
0 -63 V
1485 400 M
0 63 V
0 1381 R
0 -63 V
1813 400 M
0 63 V
0 1381 R
0 -63 V
2142 400 M
0 63 V
0 1381 R
0 -63 V
2470 400 M
0 63 V
0 1381 R
0 -63 V
LTa
500 400 M
1970 0 V
LTa
1485 400 M
0 1444 V
LTb
500 400 M
1970 0 V
0 1444 V
-1970 0 V
500 400 L
1.000 UL
LT0
2107 1731 M
263 0 V
615 403 M
16 3 V
17 3 V
16 4 V
17 3 V
17 4 V
16 3 V
17 4 V
16 4 V
17 4 V
17 5 V
16 4 V
17 5 V
17 5 V
16 6 V
17 7 V
16 7 V
17 8 V
17 9 V
16 11 V
17 12 V
16 14 V
17 16 V
17 18 V
16 20 V
17 22 V
16 25 V
17 28 V
17 30 V
16 32 V
17 34 V
16 36 V
17 38 V
17 40 V
16 41 V
17 43 V
16 44 V
17 46 V
17 47 V
16 47 V
17 48 V
17 48 V
16 47 V
17 46 V
16 44 V
17 42 V
17 39 V
16 36 V
17 33 V
16 28 V
17 23 V
17 19 V
16 14 V
17 8 V
16 4 V
17 -2 V
17 -7 V
16 -12 V
17 -18 V
16 -23 V
17 -29 V
17 -35 V
16 -40 V
17 -46 V
16 -51 V
17 -55 V
17 -57 V
16 -60 V
17 -61 V
16 -61 V
17 -60 V
17 -59 V
16 -56 V
17 -52 V
17 -49 V
16 -43 V
17 -38 V
16 -33 V
17 -29 V
17 -25 V
16 -21 V
17 -19 V
16 -15 V
17 -14 V
17 -11 V
16 -10 V
17 -9 V
16 -9 V
17 -8 V
17 -8 V
16 -7 V
17 -8 V
16 -7 V
17 -7 V
17 -7 V
16 -6 V
17 -7 V
16 -6 V
17 -7 V
17 -6 V
1.000 UL
LT1
2107 1631 M
263 0 V
662 406 M
16 3 V
16 2 V
17 3 V
16 4 V
16 3 V
16 5 V
16 5 V
17 5 V
16 7 V
16 7 V
16 9 V
17 9 V
16 11 V
16 12 V
16 14 V
17 15 V
16 17 V
16 18 V
16 20 V
17 22 V
16 23 V
16 26 V
16 27 V
16 29 V
17 31 V
16 33 V
16 35 V
16 36 V
17 38 V
16 38 V
16 39 V
16 40 V
17 39 V
16 40 V
16 39 V
16 39 V
16 38 V
17 36 V
16 35 V
16 33 V
16 32 V
17 29 V
16 26 V
16 23 V
16 20 V
17 16 V
16 13 V
16 8 V
16 4 V
17 -1 V
16 -6 V
16 -10 V
16 -14 V
16 -17 V
17 -20 V
16 -22 V
16 -25 V
16 -26 V
17 -27 V
16 -27 V
16 -28 V
16 -27 V
17 -27 V
16 -26 V
16 -26 V
16 -25 V
17 -26 V
16 -26 V
16 -27 V
16 -28 V
16 -28 V
17 -29 V
16 -31 V
16 -32 V
16 -34 V
17 -36 V
16 -36 V
16 -37 V
16 -37 V
17 -37 V
16 -36 V
16 -35 V
16 -33 V
17 -30 V
16 -28 V
16 -25 V
16 -21 V
16 -18 V
17 -14 V
16 -11 V
16 -8 V
14 -5 V
100 0 R
16 3 V
1.000 UP
1.000 UL
LT6
1456 1574 CircleF
1456 1458 CircleF
2238 1531 CircleF
stroke
grestore
end
showpage
}
\put(2057,1531){\makebox(0,0)[r]{COBE }}
\put(2057,1631){\makebox(0,0)[r]{$\alpha$=0.2}}
\put(2057,1731){\makebox(0,0)[r]{$\alpha$=0}}
\put(1485,150){\makebox(0,0){skewness}}
\put(100,1122){%
\special{ps: gsave currentpoint currentpoint translate
270 rotate neg exch neg exch translate}%
\makebox(0,0)[b]{\shortstack{Probability}}%
\special{ps: currentpoint grestore moveto}%
}
\put(2470,300){\makebox(0,0){0.3}}
\put(2142,300){\makebox(0,0){0.2}}
\put(1813,300){\makebox(0,0){0.1}}
\put(1485,300){\makebox(0,0){0}}
\put(1157,300){\makebox(0,0){-0.1}}
\put(828,300){\makebox(0,0){-0.2}}
\put(500,300){\makebox(0,0){-0.3}}
\put(450,1844){\makebox(0,0)[r]{0.4}}
\put(450,1664){\makebox(0,0)[r]{0.35}}
\put(450,1483){\makebox(0,0)[r]{0.3}}
\put(450,1303){\makebox(0,0)[r]{0.25}}
\put(450,1122){\makebox(0,0)[r]{0.2}}
\put(450,942){\makebox(0,0)[r]{0.15}}
\put(450,761){\makebox(0,0)[r]{0.1}}
\put(450,581){\makebox(0,0)[r]{0.05}}
\put(450,400){\makebox(0,0)[r]{0}}
\end{picture}

%% file: plotmsd.latex
\setlength{\unitlength}{0.1bp}
\special{!
/gnudict 120 dict def
gnudict begin
/Color false def
/Solid false def
/gnulinewidth 5.000 def
/userlinewidth gnulinewidth def
/vshift -33 def
/dl {10 mul} def
/hpt_ 31.5 def
/vpt_ 31.5 def
/hpt hpt_ def
/vpt vpt_ def
/M {moveto} bind def
/L {lineto} bind def
/R {rmoveto} bind def
/V {rlineto} bind def
/vpt2 vpt 2 mul def
/hpt2 hpt 2 mul def
/Lshow { currentpoint stroke M
  0 vshift R show } def
/Rshow { currentpoint stroke M
  dup stringwidth pop neg vshift R show } def
/Cshow { currentpoint stroke M
  dup stringwidth pop -2 div vshift R show } def
/UP { dup vpt_ mul /vpt exch def hpt_ mul /hpt exch def
  /hpt2 hpt 2 mul def /vpt2 vpt 2 mul def } def
/DL { Color {setrgbcolor Solid {pop []} if 0 setdash }
 {pop pop pop Solid {pop []} if 0 setdash} ifelse } def
/BL { stroke gnulinewidth 2 mul setlinewidth } def
/AL { stroke gnulinewidth 2 div setlinewidth } def
/UL { gnulinewidth mul /userlinewidth exch def } def
/PL { stroke userlinewidth setlinewidth } def
/LTb { BL [] 0 0 0 DL } def
/LTa { AL [1 dl 2 dl] 0 setdash 0 0 0 setrgbcolor } def
/LT0 { PL [] 0 1 0 DL } def
/LT1 { PL [4 dl 2 dl] 0 0 1 DL } def
/LT2 { PL [2 dl 3 dl] 1 0 0 DL } def
/LT3 { PL [1 dl 1.5 dl] 1 0 1 DL } def
/LT4 { PL [5 dl 2 dl 1 dl 2 dl] 0 1 1 DL } def
/LT5 { PL [4 dl 3 dl 1 dl 3 dl] 1 1 0 DL } def
/LT6 { PL [2 dl 2 dl 2 dl 4 dl] 0 0 0 DL } def
/LT7 { PL [2 dl 2 dl 2 dl 2 dl 2 dl 4 dl] 1 0.3 0 DL } def
/LT8 { PL [2 dl 2 dl 2 dl 2 dl 2 dl 2 dl 2 dl 4 dl] 0.5 0.5 0.5 DL } def
/Pnt { stroke [] 0 setdash
   gsave 1 setlinecap M 0 0 V stroke grestore } def
/Dia { stroke [] 0 setdash 2 copy vpt add M
  hpt neg vpt neg V hpt vpt neg V
  hpt vpt V hpt neg vpt V closepath stroke
  Pnt } def
/Pls { stroke [] 0 setdash vpt sub M 0 vpt2 V
  currentpoint stroke M
  hpt neg vpt neg R hpt2 0 V stroke
  } def
/Box { stroke [] 0 setdash 2 copy exch hpt sub exch vpt add M
  0 vpt2 neg V hpt2 0 V 0 vpt2 V
  hpt2 neg 0 V closepath stroke
  Pnt } def
/Crs { stroke [] 0 setdash exch hpt sub exch vpt add M
  hpt2 vpt2 neg V currentpoint stroke M
  hpt2 neg 0 R hpt2 vpt2 V stroke } def
/TriU { stroke [] 0 setdash 2 copy vpt 1.12 mul add M
  hpt neg vpt -1.62 mul V
  hpt 2 mul 0 V
  hpt neg vpt 1.62 mul V closepath stroke
  Pnt  } def
/Star { 2 copy Pls Crs } def
/BoxF { stroke [] 0 setdash exch hpt sub exch vpt add M
  0 vpt2 neg V  hpt2 0 V  0 vpt2 V
  hpt2 neg 0 V  closepath fill } def
/TriUF { stroke [] 0 setdash vpt 1.12 mul add M
  hpt neg vpt -1.62 mul V
  hpt 2 mul 0 V
  hpt neg vpt 1.62 mul V closepath fill } def
/TriD { stroke [] 0 setdash 2 copy vpt 1.12 mul sub M
  hpt neg vpt 1.62 mul V
  hpt 2 mul 0 V
  hpt neg vpt -1.62 mul V closepath stroke
  Pnt  } def
/TriDF { stroke [] 0 setdash vpt 1.12 mul sub M
  hpt neg vpt 1.62 mul V
  hpt 2 mul 0 V
  hpt neg vpt -1.62 mul V closepath fill} def
/DiaF { stroke [] 0 setdash vpt add M
  hpt neg vpt neg V hpt vpt neg V
  hpt vpt V hpt neg vpt V closepath fill } def
/Pent { stroke [] 0 setdash 2 copy gsave
  translate 0 hpt M 4 {72 rotate 0 hpt L} repeat
  closepath stroke grestore Pnt } def
/PentF { stroke [] 0 setdash gsave
  translate 0 hpt M 4 {72 rotate 0 hpt L} repeat
  closepath fill grestore } def
/Circle { stroke [] 0 setdash 2 copy
  hpt 0 360 arc stroke Pnt } def
/CircleF { stroke [] 0 setdash hpt 0 360 arc fill } def
/C0 { BL [] 0 setdash 2 copy moveto vpt 90 450  arc } bind def
/C1 { BL [] 0 setdash 2 copy        moveto
       2 copy  vpt 0 90 arc closepath fill
               vpt 0 360 arc closepath } bind def
/C2 { BL [] 0 setdash 2 copy moveto
       2 copy  vpt 90 180 arc closepath fill
               vpt 0 360 arc closepath } bind def
/C3 { BL [] 0 setdash 2 copy moveto
       2 copy  vpt 0 180 arc closepath fill
               vpt 0 360 arc closepath } bind def
/C4 { BL [] 0 setdash 2 copy moveto
       2 copy  vpt 180 270 arc closepath fill
               vpt 0 360 arc closepath } bind def
/C5 { BL [] 0 setdash 2 copy moveto
       2 copy  vpt 0 90 arc
       2 copy moveto
       2 copy  vpt 180 270 arc closepath fill
               vpt 0 360 arc } bind def
/C6 { BL [] 0 setdash 2 copy moveto
      2 copy  vpt 90 270 arc closepath fill
              vpt 0 360 arc closepath } bind def
/C7 { BL [] 0 setdash 2 copy moveto
      2 copy  vpt 0 270 arc closepath fill
              vpt 0 360 arc closepath } bind def
/C8 { BL [] 0 setdash 2 copy moveto
      2 copy vpt 270 360 arc closepath fill
              vpt 0 360 arc closepath } bind def
/C9 { BL [] 0 setdash 2 copy moveto
      2 copy  vpt 270 450 arc closepath fill
              vpt 0 360 arc closepath } bind def
/C10 { BL [] 0 setdash 2 copy 2 copy moveto vpt 270 360 arc closepath fill
       2 copy moveto
       2 copy vpt 90 180 arc closepath fill
               vpt 0 360 arc closepath } bind def
/C11 { BL [] 0 setdash 2 copy moveto
       2 copy  vpt 0 180 arc closepath fill
       2 copy moveto
       2 copy  vpt 270 360 arc closepath fill
               vpt 0 360 arc closepath } bind def
/C12 { BL [] 0 setdash 2 copy moveto
       2 copy  vpt 180 360 arc closepath fill
               vpt 0 360 arc closepath } bind def
/C13 { BL [] 0 setdash  2 copy moveto
       2 copy  vpt 0 90 arc closepath fill
       2 copy moveto
       2 copy  vpt 180 360 arc closepath fill
               vpt 0 360 arc closepath } bind def
/C14 { BL [] 0 setdash 2 copy moveto
       2 copy  vpt 90 360 arc closepath fill
               vpt 0 360 arc } bind def
/C15 { BL [] 0 setdash 2 copy vpt 0 360 arc closepath fill
               vpt 0 360 arc closepath } bind def
/Rec   { newpath 4 2 roll moveto 1 index 0 rlineto 0 exch rlineto
       neg 0 rlineto closepath } bind def
/Square { dup Rec } bind def
/Bsquare { vpt sub exch vpt sub exch vpt2 Square } bind def
/S0 { BL [] 0 setdash 2 copy moveto 0 vpt rlineto BL Bsquare } bind def
/S1 { BL [] 0 setdash 2 copy vpt Square fill Bsquare } bind def
/S2 { BL [] 0 setdash 2 copy exch vpt sub exch vpt Square fill Bsquare } bind def
/S3 { BL [] 0 setdash 2 copy exch vpt sub exch vpt2 vpt Rec fill Bsquare } bind def
/S4 { BL [] 0 setdash 2 copy exch vpt sub exch vpt sub vpt Square fill Bsquare } bind def
/S5 { BL [] 0 setdash 2 copy 2 copy vpt Square fill
       exch vpt sub exch vpt sub vpt Square fill Bsquare } bind def
/S6 { BL [] 0 setdash 2 copy exch vpt sub exch vpt sub vpt vpt2 Rec fill Bsquare } bind def
/S7 { BL [] 0 setdash 2 copy exch vpt sub exch vpt sub vpt vpt2 Rec fill
       2 copy vpt Square fill
       Bsquare } bind def
/S8 { BL [] 0 setdash 2 copy vpt sub vpt Square fill Bsquare } bind def
/S9 { BL [] 0 setdash 2 copy vpt sub vpt vpt2 Rec fill Bsquare } bind def
/S10 { BL [] 0 setdash 2 copy vpt sub vpt Square fill 2 copy exch vpt sub exch vpt Square fill
       Bsquare } bind def
/S11 { BL [] 0 setdash 2 copy vpt sub vpt Square fill 2 copy exch vpt sub exch vpt2 vpt Rec fill
       Bsquare } bind def
/S12 { BL [] 0 setdash 2 copy exch vpt sub exch vpt sub vpt2 vpt Rec fill Bsquare } bind def
/S13 { BL [] 0 setdash 2 copy exch vpt sub exch vpt sub vpt2 vpt Rec fill
       2 copy vpt Square fill Bsquare } bind def
/S14 { BL [] 0 setdash 2 copy exch vpt sub exch vpt sub vpt2 vpt Rec fill
       2 copy exch vpt sub exch vpt Square fill Bsquare } bind def
/S15 { BL [] 0 setdash 2 copy Bsquare fill Bsquare } bind def
/D0 { gsave translate 45 rotate 0 0 S0 stroke grestore } bind def
/D1 { gsave translate 45 rotate 0 0 S1 stroke grestore } bind def
/D2 { gsave translate 45 rotate 0 0 S2 stroke grestore } bind def
/D3 { gsave translate 45 rotate 0 0 S3 stroke grestore } bind def
/D4 { gsave translate 45 rotate 0 0 S4 stroke grestore } bind def
/D5 { gsave translate 45 rotate 0 0 S5 stroke grestore } bind def
/D6 { gsave translate 45 rotate 0 0 S6 stroke grestore } bind def
/D7 { gsave translate 45 rotate 0 0 S7 stroke grestore } bind def
/D8 { gsave translate 45 rotate 0 0 S8 stroke grestore } bind def
/D9 { gsave translate 45 rotate 0 0 S9 stroke grestore } bind def
/D10 { gsave translate 45 rotate 0 0 S10 stroke grestore } bind def
/D11 { gsave translate 45 rotate 0 0 S11 stroke grestore } bind def
/D12 { gsave translate 45 rotate 0 0 S12 stroke grestore } bind def
/D13 { gsave translate 45 rotate 0 0 S13 stroke grestore } bind def
/D14 { gsave translate 45 rotate 0 0 S14 stroke grestore } bind def
/D15 { gsave translate 45 rotate 0 0 S15 stroke grestore } bind def
/DiaE { stroke [] 0 setdash vpt add M
  hpt neg vpt neg V hpt vpt neg V
  hpt vpt V hpt neg vpt V closepath stroke } def
/BoxE { stroke [] 0 setdash exch hpt sub exch vpt add M
  0 vpt2 neg V hpt2 0 V 0 vpt2 V
  hpt2 neg 0 V closepath stroke } def
/TriUE { stroke [] 0 setdash vpt 1.12 mul add M
  hpt neg vpt -1.62 mul V
  hpt 2 mul 0 V
  hpt neg vpt 1.62 mul V closepath stroke } def
/TriDE { stroke [] 0 setdash vpt 1.12 mul sub M
  hpt neg vpt 1.62 mul V
  hpt 2 mul 0 V
  hpt neg vpt -1.62 mul V closepath stroke } def
/PentE { stroke [] 0 setdash gsave
  translate 0 hpt M 4 {72 rotate 0 hpt L} repeat
  closepath stroke grestore } def
/CircE { stroke [] 0 setdash 
  hpt 0 360 arc stroke } def
/BoxFill { gsave Rec 1 setgray fill grestore } def
end
}
\begin{picture}(2520,1944)(0,0)
\special{"
gnudict begin
gsave
0 0 translate
0.100 0.100 scale
0 setgray
newpath
LTb
500 400 M
63 0 V
1907 0 R
-63 0 V
500 544 M
63 0 V
1907 0 R
-63 0 V
500 689 M
63 0 V
1907 0 R
-63 0 V
500 833 M
63 0 V
1907 0 R
-63 0 V
500 978 M
63 0 V
1907 0 R
-63 0 V
500 1122 M
63 0 V
1907 0 R
-63 0 V
500 1266 M
63 0 V
1907 0 R
-63 0 V
500 1411 M
63 0 V
1907 0 R
-63 0 V
500 1555 M
63 0 V
1907 0 R
-63 0 V
500 1700 M
63 0 V
1907 0 R
-63 0 V
500 1844 M
63 0 V
1907 0 R
-63 0 V
500 400 M
0 63 V
0 1381 R
0 -63 V
679 400 M
0 63 V
0 1381 R
0 -63 V
858 400 M
0 63 V
0 1381 R
0 -63 V
1037 400 M
0 63 V
0 1381 R
0 -63 V
1216 400 M
0 63 V
0 1381 R
0 -63 V
1395 400 M
0 63 V
0 1381 R
0 -63 V
1575 400 M
0 63 V
0 1381 R
0 -63 V
1754 400 M
0 63 V
0 1381 R
0 -63 V
1933 400 M
0 63 V
0 1381 R
0 -63 V
2112 400 M
0 63 V
0 1381 R
0 -63 V
2291 400 M
0 63 V
0 1381 R
0 -63 V
2470 400 M
0 63 V
0 1381 R
0 -63 V
LTb
500 400 M
1970 0 V
0 1444 V
-1970 0 V
500 400 L
1.000 UL
LT0
2107 1731 M
263 0 V
618 542 M
13 87 V
13 87 V
13 85 V
13 82 V
13 79 V
13 74 V
13 70 V
13 64 V
13 57 V
13 49 V
13 41 V
13 32 V
13 21 V
14 12 V
13 3 V
13 -5 V
13 -12 V
13 -18 V
13 -22 V
13 -27 V
13 -29 V
13 -32 V
13 -33 V
13 -32 V
13 -32 V
13 -30 V
13 -29 V
13 -27 V
13 -25 V
13 -24 V
13 -23 V
13 -22 V
13 -20 V
13 -20 V
13 -18 V
13 -17 V
13 -17 V
13 -15 V
13 -15 V
13 -14 V
14 -14 V
13 -13 V
13 -12 V
13 -13 V
13 -12 V
13 -12 V
13 -12 V
13 -12 V
13 -13 V
13 -12 V
13 -13 V
13 -14 V
13 -13 V
13 -13 V
13 -14 V
13 -13 V
13 -14 V
13 -13 V
13 -13 V
13 -13 V
13 -12 V
13 -12 V
13 -11 V
13 -11 V
13 -10 V
13 -10 V
14 -8 V
13 -9 V
13 -7 V
13 -7 V
13 -6 V
13 -5 V
13 -4 V
13 -3 V
13 -3 V
13 -1 V
13 -1 V
13 0 V
13 0 V
13 1 V
13 1 V
13 1 V
13 1 V
13 0 V
13 1 V
13 -1 V
13 0 V
13 -2 V
13 -2 V
13 -2 V
13 -4 V
13 -3 V
14 -4 V
13 -4 V
13 -5 V
13 -5 V
13 -5 V
13 -5 V
13 -5 V
1.000 UL
LT1
2107 1631 M
263 0 V
682 463 M
17 32 V
16 32 V
16 33 V
17 34 V
16 36 V
16 37 V
16 40 V
17 42 V
16 45 V
16 48 V
17 53 V
16 56 V
16 60 V
16 64 V
17 66 V
16 66 V
16 66 V
17 62 V
16 58 V
16 53 V
16 45 V
17 36 V
16 26 V
16 13 V
17 0 V
16 -14 V
16 -27 V
16 -38 V
17 -47 V
16 -54 V
16 -59 V
17 -63 V
16 -65 V
16 -65 V
16 -63 V
17 -59 V
16 -55 V
16 -47 V
16 -41 V
17 -34 V
16 -29 V
16 -24 V
17 -20 V
16 -17 V
16 -13 V
16 -12 V
17 -9 V
16 -9 V
16 -9 V
17 -9 V
16 -9 V
16 -10 V
16 -11 V
17 -11 V
16 -11 V
16 -11 V
17 -12 V
16 -11 V
16 -11 V
16 -11 V
17 -11 V
16 -10 V
16 -10 V
17 -9 V
16 -9 V
16 -8 V
16 -8 V
17 -7 V
16 -7 V
16 -7 V
17 -6 V
16 -6 V
16 -6 V
16 -6 V
17 -5 V
16 -5 V
16 -5 V
17 -4 V
16 -4 V
16 -4 V
16 -4 V
17 -3 V
16 -3 V
16 -3 V
16 -3 V
17 -2 V
16 -3 V
16 -2 V
17 -1 V
16 -2 V
16 -1 V
16 -1 V
17 -1 V
16 -1 V
16 -1 V
17 -1 V
16 -1 V
16 -1 V
16 0 V
1.000 UP
1.000 UL
LT6
1086 895 CircleF
1086 1566 CircleF
2238 1531 CircleF
stroke
grestore
end
showpage
}
\put(2057,1531){\makebox(0,0)[r]{COBE}}
\put(2057,1631){\makebox(0,0)[r]{$\alpha$=0.2}}
\put(2057,1731){\makebox(0,0)[r]{$\alpha$=0}}
\put(1485,150){\makebox(0,0){MSD}}
\put(100,1122){%
\special{ps: gsave currentpoint currentpoint translate
270 rotate neg exch neg exch translate}%
\makebox(0,0)[b]{\shortstack{Probability}}%
\special{ps: currentpoint grestore moveto}%
}
\put(2470,300){\makebox(0,0){22}}
\put(2291,300){\makebox(0,0){20}}
\put(2112,300){\makebox(0,0){18}}
\put(1933,300){\makebox(0,0){16}}
\put(1754,300){\makebox(0,0){14}}
\put(1575,300){\makebox(0,0){12}}
\put(1395,300){\makebox(0,0){10}}
\put(1216,300){\makebox(0,0){8}}
\put(1037,300){\makebox(0,0){6}}
\put(858,300){\makebox(0,0){4}}
\put(679,300){\makebox(0,0){2}}
\put(500,300){\makebox(0,0){0}}
\put(450,1844){\makebox(0,0)[r]{0.5}}
\put(450,1700){\makebox(0,0)[r]{0.45}}
\put(450,1555){\makebox(0,0)[r]{0.4}}
\put(450,1411){\makebox(0,0)[r]{0.35}}
\put(450,1266){\makebox(0,0)[r]{0.3}}
\put(450,1122){\makebox(0,0)[r]{0.25}}
\put(450,978){\makebox(0,0)[r]{0.2}}
\put(450,833){\makebox(0,0)[r]{0.15}}
\put(450,689){\makebox(0,0)[r]{0.1}}
\put(450,544){\makebox(0,0)[r]{0.05}}
\put(450,400){\makebox(0,0)[r]{0}}
\end{picture}

%% file: plotnorth.latex
\setlength{\unitlength}{0.1bp}
\special{!
/gnudict 120 dict def
gnudict begin
/Color false def
/Solid false def
/gnulinewidth 5.000 def
/userlinewidth gnulinewidth def
/vshift -33 def
/dl {10 mul} def
/hpt_ 31.5 def
/vpt_ 31.5 def
/hpt hpt_ def
/vpt vpt_ def
/M {moveto} bind def
/L {lineto} bind def
/R {rmoveto} bind def
/V {rlineto} bind def
/vpt2 vpt 2 mul def
/hpt2 hpt 2 mul def
/Lshow { currentpoint stroke M
  0 vshift R show } def
/Rshow { currentpoint stroke M
  dup stringwidth pop neg vshift R show } def
/Cshow { currentpoint stroke M
  dup stringwidth pop -2 div vshift R show } def
/UP { dup vpt_ mul /vpt exch def hpt_ mul /hpt exch def
  /hpt2 hpt 2 mul def /vpt2 vpt 2 mul def } def
/DL { Color {setrgbcolor Solid {pop []} if 0 setdash }
 {pop pop pop Solid {pop []} if 0 setdash} ifelse } def
/BL { stroke gnulinewidth 2 mul setlinewidth } def
/AL { stroke gnulinewidth 2 div setlinewidth } def
/UL { gnulinewidth mul /userlinewidth exch def } def
/PL { stroke userlinewidth setlinewidth } def
/LTb { BL [] 0 0 0 DL } def
/LTa { AL [1 dl 2 dl] 0 setdash 0 0 0 setrgbcolor } def
/LT0 { PL [] 0 1 0 DL } def
/LT1 { PL [4 dl 2 dl] 0 0 1 DL } def
/LT2 { PL [2 dl 3 dl] 1 0 0 DL } def
/LT3 { PL [1 dl 1.5 dl] 1 0 1 DL } def
/LT4 { PL [5 dl 2 dl 1 dl 2 dl] 0 1 1 DL } def
/LT5 { PL [4 dl 3 dl 1 dl 3 dl] 1 1 0 DL } def
/LT6 { PL [2 dl 2 dl 2 dl 4 dl] 0 0 0 DL } def
/LT7 { PL [2 dl 2 dl 2 dl 2 dl 2 dl 4 dl] 1 0.3 0 DL } def
/LT8 { PL [2 dl 2 dl 2 dl 2 dl 2 dl 2 dl 2 dl 4 dl] 0.5 0.5 0.5 DL } def
/Pnt { stroke [] 0 setdash
   gsave 1 setlinecap M 0 0 V stroke grestore } def
/Dia { stroke [] 0 setdash 2 copy vpt add M
  hpt neg vpt neg V hpt vpt neg V
  hpt vpt V hpt neg vpt V closepath stroke
  Pnt } def
/Pls { stroke [] 0 setdash vpt sub M 0 vpt2 V
  currentpoint stroke M
  hpt neg vpt neg R hpt2 0 V stroke
  } def
/Box { stroke [] 0 setdash 2 copy exch hpt sub exch vpt add M
  0 vpt2 neg V hpt2 0 V 0 vpt2 V
  hpt2 neg 0 V closepath stroke
  Pnt } def
/Crs { stroke [] 0 setdash exch hpt sub exch vpt add M
  hpt2 vpt2 neg V currentpoint stroke M
  hpt2 neg 0 R hpt2 vpt2 V stroke } def
/TriU { stroke [] 0 setdash 2 copy vpt 1.12 mul add M
  hpt neg vpt -1.62 mul V
  hpt 2 mul 0 V
  hpt neg vpt 1.62 mul V closepath stroke
  Pnt  } def
/Star { 2 copy Pls Crs } def
/BoxF { stroke [] 0 setdash exch hpt sub exch vpt add M
  0 vpt2 neg V  hpt2 0 V  0 vpt2 V
  hpt2 neg 0 V  closepath fill } def
/TriUF { stroke [] 0 setdash vpt 1.12 mul add M
  hpt neg vpt -1.62 mul V
  hpt 2 mul 0 V
  hpt neg vpt 1.62 mul V closepath fill } def
/TriD { stroke [] 0 setdash 2 copy vpt 1.12 mul sub M
  hpt neg vpt 1.62 mul V
  hpt 2 mul 0 V
  hpt neg vpt -1.62 mul V closepath stroke
  Pnt  } def
/TriDF { stroke [] 0 setdash vpt 1.12 mul sub M
  hpt neg vpt 1.62 mul V
  hpt 2 mul 0 V
  hpt neg vpt -1.62 mul V closepath fill} def
/DiaF { stroke [] 0 setdash vpt add M
  hpt neg vpt neg V hpt vpt neg V
  hpt vpt V hpt neg vpt V closepath fill } def
/Pent { stroke [] 0 setdash 2 copy gsave
  translate 0 hpt M 4 {72 rotate 0 hpt L} repeat
  closepath stroke grestore Pnt } def
/PentF { stroke [] 0 setdash gsave
  translate 0 hpt M 4 {72 rotate 0 hpt L} repeat
  closepath fill grestore } def
/Circle { stroke [] 0 setdash 2 copy
  hpt 0 360 arc stroke Pnt } def
/CircleF { stroke [] 0 setdash hpt 0 360 arc fill } def
/C0 { BL [] 0 setdash 2 copy moveto vpt 90 450  arc } bind def
/C1 { BL [] 0 setdash 2 copy        moveto
       2 copy  vpt 0 90 arc closepath fill
               vpt 0 360 arc closepath } bind def
/C2 { BL [] 0 setdash 2 copy moveto
       2 copy  vpt 90 180 arc closepath fill
               vpt 0 360 arc closepath } bind def
/C3 { BL [] 0 setdash 2 copy moveto
       2 copy  vpt 0 180 arc closepath fill
               vpt 0 360 arc closepath } bind def
/C4 { BL [] 0 setdash 2 copy moveto
       2 copy  vpt 180 270 arc closepath fill
               vpt 0 360 arc closepath } bind def
/C5 { BL [] 0 setdash 2 copy moveto
       2 copy  vpt 0 90 arc
       2 copy moveto
       2 copy  vpt 180 270 arc closepath fill
               vpt 0 360 arc } bind def
/C6 { BL [] 0 setdash 2 copy moveto
      2 copy  vpt 90 270 arc closepath fill
              vpt 0 360 arc closepath } bind def
/C7 { BL [] 0 setdash 2 copy moveto
      2 copy  vpt 0 270 arc closepath fill
              vpt 0 360 arc closepath } bind def
/C8 { BL [] 0 setdash 2 copy moveto
      2 copy vpt 270 360 arc closepath fill
              vpt 0 360 arc closepath } bind def
/C9 { BL [] 0 setdash 2 copy moveto
      2 copy  vpt 270 450 arc closepath fill
              vpt 0 360 arc closepath } bind def
/C10 { BL [] 0 setdash 2 copy 2 copy moveto vpt 270 360 arc closepath fill
       2 copy moveto
       2 copy vpt 90 180 arc closepath fill
               vpt 0 360 arc closepath } bind def
/C11 { BL [] 0 setdash 2 copy moveto
       2 copy  vpt 0 180 arc closepath fill
       2 copy moveto
       2 copy  vpt 270 360 arc closepath fill
               vpt 0 360 arc closepath } bind def
/C12 { BL [] 0 setdash 2 copy moveto
       2 copy  vpt 180 360 arc closepath fill
               vpt 0 360 arc closepath } bind def
/C13 { BL [] 0 setdash  2 copy moveto
       2 copy  vpt 0 90 arc closepath fill
       2 copy moveto
       2 copy  vpt 180 360 arc closepath fill
               vpt 0 360 arc closepath } bind def
/C14 { BL [] 0 setdash 2 copy moveto
       2 copy  vpt 90 360 arc closepath fill
               vpt 0 360 arc } bind def
/C15 { BL [] 0 setdash 2 copy vpt 0 360 arc closepath fill
               vpt 0 360 arc closepath } bind def
/Rec   { newpath 4 2 roll moveto 1 index 0 rlineto 0 exch rlineto
       neg 0 rlineto closepath } bind def
/Square { dup Rec } bind def
/Bsquare { vpt sub exch vpt sub exch vpt2 Square } bind def
/S0 { BL [] 0 setdash 2 copy moveto 0 vpt rlineto BL Bsquare } bind def
/S1 { BL [] 0 setdash 2 copy vpt Square fill Bsquare } bind def
/S2 { BL [] 0 setdash 2 copy exch vpt sub exch vpt Square fill Bsquare } bind def
/S3 { BL [] 0 setdash 2 copy exch vpt sub exch vpt2 vpt Rec fill Bsquare } bind def
/S4 { BL [] 0 setdash 2 copy exch vpt sub exch vpt sub vpt Square fill Bsquare } bind def
/S5 { BL [] 0 setdash 2 copy 2 copy vpt Square fill
       exch vpt sub exch vpt sub vpt Square fill Bsquare } bind def
/S6 { BL [] 0 setdash 2 copy exch vpt sub exch vpt sub vpt vpt2 Rec fill Bsquare } bind def
/S7 { BL [] 0 setdash 2 copy exch vpt sub exch vpt sub vpt vpt2 Rec fill
       2 copy vpt Square fill
       Bsquare } bind def
/S8 { BL [] 0 setdash 2 copy vpt sub vpt Square fill Bsquare } bind def
/S9 { BL [] 0 setdash 2 copy vpt sub vpt vpt2 Rec fill Bsquare } bind def
/S10 { BL [] 0 setdash 2 copy vpt sub vpt Square fill 2 copy exch vpt sub exch vpt Square fill
       Bsquare } bind def
/S11 { BL [] 0 setdash 2 copy vpt sub vpt Square fill 2 copy exch vpt sub exch vpt2 vpt Rec fill
       Bsquare } bind def
/S12 { BL [] 0 setdash 2 copy exch vpt sub exch vpt sub vpt2 vpt Rec fill Bsquare } bind def
/S13 { BL [] 0 setdash 2 copy exch vpt sub exch vpt sub vpt2 vpt Rec fill
       2 copy vpt Square fill Bsquare } bind def
/S14 { BL [] 0 setdash 2 copy exch vpt sub exch vpt sub vpt2 vpt Rec fill
       2 copy exch vpt sub exch vpt Square fill Bsquare } bind def
/S15 { BL [] 0 setdash 2 copy Bsquare fill Bsquare } bind def
/D0 { gsave translate 45 rotate 0 0 S0 stroke grestore } bind def
/D1 { gsave translate 45 rotate 0 0 S1 stroke grestore } bind def
/D2 { gsave translate 45 rotate 0 0 S2 stroke grestore } bind def
/D3 { gsave translate 45 rotate 0 0 S3 stroke grestore } bind def
/D4 { gsave translate 45 rotate 0 0 S4 stroke grestore } bind def
/D5 { gsave translate 45 rotate 0 0 S5 stroke grestore } bind def
/D6 { gsave translate 45 rotate 0 0 S6 stroke grestore } bind def
/D7 { gsave translate 45 rotate 0 0 S7 stroke grestore } bind def
/D8 { gsave translate 45 rotate 0 0 S8 stroke grestore } bind def
/D9 { gsave translate 45 rotate 0 0 S9 stroke grestore } bind def
/D10 { gsave translate 45 rotate 0 0 S10 stroke grestore } bind def
/D11 { gsave translate 45 rotate 0 0 S11 stroke grestore } bind def
/D12 { gsave translate 45 rotate 0 0 S12 stroke grestore } bind def
/D13 { gsave translate 45 rotate 0 0 S13 stroke grestore } bind def
/D14 { gsave translate 45 rotate 0 0 S14 stroke grestore } bind def
/D15 { gsave translate 45 rotate 0 0 S15 stroke grestore } bind def
/DiaE { stroke [] 0 setdash vpt add M
  hpt neg vpt neg V hpt vpt neg V
  hpt vpt V hpt neg vpt V closepath stroke } def
/BoxE { stroke [] 0 setdash exch hpt sub exch vpt add M
  0 vpt2 neg V hpt2 0 V 0 vpt2 V
  hpt2 neg 0 V closepath stroke } def
/TriUE { stroke [] 0 setdash vpt 1.12 mul add M
  hpt neg vpt -1.62 mul V
  hpt 2 mul 0 V
  hpt neg vpt 1.62 mul V closepath stroke } def
/TriDE { stroke [] 0 setdash vpt 1.12 mul sub M
  hpt neg vpt 1.62 mul V
  hpt 2 mul 0 V
  hpt neg vpt -1.62 mul V closepath stroke } def
/PentE { stroke [] 0 setdash gsave
  translate 0 hpt M 4 {72 rotate 0 hpt L} repeat
  closepath stroke grestore } def
/CircE { stroke [] 0 setdash 
  hpt 0 360 arc stroke } def
/BoxFill { gsave Rec 1 setgray fill grestore } def
end
}
\begin{picture}(2520,2160)(0,0)
\special{"
gnudict begin
gsave
0 0 translate
0.100 0.100 scale
0 setgray
newpath
LTb
450 400 M
63 0 V
1957 0 R
-63 0 V
450 652 M
63 0 V
1957 0 R
-63 0 V
450 903 M
63 0 V
1957 0 R
-63 0 V
450 1155 M
63 0 V
1957 0 R
-63 0 V
450 1407 M
63 0 V
1957 0 R
-63 0 V
450 1658 M
63 0 V
1957 0 R
-63 0 V
450 1910 M
63 0 V
1957 0 R
-63 0 V
450 400 M
0 63 V
0 1447 R
0 -63 V
854 400 M
0 63 V
0 1447 R
0 -63 V
1258 400 M
0 63 V
0 1447 R
0 -63 V
1662 400 M
0 63 V
0 1447 R
0 -63 V
2066 400 M
0 63 V
0 1447 R
0 -63 V
2470 400 M
0 63 V
0 1447 R
0 -63 V
LTb
450 400 M
2020 0 V
0 1510 V
-2020 0 V
450 400 L
1.000 UL
LT0
2107 1797 M
263 0 V
450 903 M
12 -2 V
12 -3 V
13 0 V
12 0 V
12 4 V
12 6 V
13 10 V
12 14 V
12 18 V
12 23 V
13 26 V
12 29 V
12 31 V
12 31 V
13 31 V
12 30 V
12 29 V
12 26 V
13 25 V
12 25 V
12 24 V
12 24 V
13 26 V
12 28 V
12 30 V
12 32 V
13 33 V
12 33 V
12 31 V
12 27 V
13 23 V
12 16 V
12 9 V
12 0 V
12 -7 V
13 -14 V
12 -19 V
12 -22 V
12 -23 V
13 -24 V
12 -23 V
12 -21 V
12 -18 V
13 -17 V
12 -18 V
12 -20 V
12 -22 V
13 -26 V
12 -31 V
12 -37 V
12 -43 V
13 -48 V
12 -51 V
12 -53 V
12 -53 V
13 -51 V
12 -48 V
12 -43 V
12 -38 V
13 -33 V
12 -29 V
12 -24 V
12 -22 V
13 -19 V
12 -17 V
12 -16 V
12 -15 V
12 -14 V
13 -13 V
12 -13 V
12 -12 V
12 -11 V
13 -11 V
12 -10 V
12 -9 V
12 -9 V
13 -9 V
12 -8 V
12 -7 V
12 -7 V
13 -6 V
12 -6 V
12 -6 V
12 -5 V
13 -5 V
12 -5 V
12 -4 V
12 -4 V
13 -4 V
12 -3 V
12 -4 V
12 -3 V
13 -3 V
12 -2 V
12 -3 V
12 -3 V
13 -2 V
12 -3 V
12 -2 V
1.000 UL
LT1
2107 1697 M
263 0 V
450 903 M
16 0 V
17 1 V
16 2 V
16 5 V
17 7 V
16 11 V
16 16 V
17 19 V
16 22 V
16 25 V
17 26 V
16 27 V
16 27 V
17 25 V
16 21 V
16 15 V
16 6 V
17 -4 V
16 -16 V
16 -29 V
17 -39 V
16 -45 V
16 -49 V
17 -50 V
16 -48 V
16 -43 V
17 -39 V
16 -33 V
16 -28 V
17 -21 V
16 -15 V
16 -8 V
17 -4 V
16 -2 V
16 -2 V
17 -7 V
16 -12 V
16 -20 V
17 -27 V
16 -30 V
16 -31 V
17 -29 V
16 -23 V
16 -16 V
17 -9 V
16 -2 V
16 2 V
17 5 V
16 6 V
16 5 V
16 5 V
17 3 V
16 2 V
16 2 V
17 0 V
16 0 V
16 -2 V
17 -2 V
16 -2 V
16 -4 V
17 -3 V
16 -5 V
16 -4 V
17 -5 V
16 -6 V
16 -5 V
17 -6 V
16 -5 V
16 -6 V
17 -5 V
16 -6 V
16 -5 V
17 -5 V
16 -5 V
16 -4 V
17 -4 V
16 -3 V
16 -3 V
17 -3 V
16 -3 V
16 -2 V
17 -1 V
16 -2 V
15 -1 V
190 0 R
7 0 V
16 1 V
17 1 V
16 1 V
1.000 UL
LT2
2107 1597 M
263 0 V
450 903 M
31 -15 V
30 -9 V
31 3 V
30 20 V
31 29 V
31 23 V
30 2 V
31 -23 V
30 -34 V
31 -29 V
31 -14 V
30 -4 V
31 1 V
30 1 V
31 0 V
31 3 V
30 7 V
31 10 V
31 6 V
30 -7 V
31 -20 V
30 -22 V
31 -11 V
31 11 V
30 27 V
31 33 V
30 28 V
31 19 V
31 12 V
30 4 V
31 -1 V
30 -7 V
31 -10 V
31 -13 V
30 -16 V
31 -16 V
30 -16 V
31 -16 V
31 -13 V
30 -10 V
31 -7 V
30 -5 V
31 -1 V
31 0 V
30 3 V
31 3 V
30 4 V
31 5 V
31 4 V
30 4 V
31 4 V
31 1 V
30 1 V
31 -2 V
30 -4 V
31 -5 V
31 -7 V
30 -8 V
31 -9 V
30 -9 V
31 -10 V
31 -10 V
30 -11 V
31 -10 V
30 -9 V
31 -10 V
1.000 UL
LT3
2107 1497 M
263 0 V
450 903 M
31 -73 V
30 -56 V
31 -21 V
30 28 V
31 65 V
31 70 V
30 46 V
31 7 V
30 -29 V
31 -60 V
31 -80 V
30 -72 V
31 -31 V
30 35 V
31 76 V
31 72 V
30 22 V
31 -34 V
31 -63 V
30 -64 V
31 -43 V
30 -26 V
31 -15 V
31 -9 V
30 -5 V
31 -3 V
30 -3 V
31 -2 V
31 -1 V
30 -2 V
31 -1 V
30 -1 V
31 0 V
31 -1 V
30 0 V
31 1 V
30 0 V
31 1 V
31 2 V
30 1 V
31 2 V
30 3 V
31 2 V
31 3 V
30 3 V
31 3 V
30 3 V
31 3 V
31 4 V
30 3 V
31 3 V
31 2 V
30 3 V
31 3 V
30 2 V
31 2 V
31 3 V
30 4 V
31 3 V
30 5 V
31 5 V
31 7 V
30 8 V
31 9 V
30 11 V
31 12 V
stroke
grestore
end
showpage
}
\put(2057,1497){\makebox(0,0)[r]{Kurtosis}}
\put(2057,1597){\makebox(0,0)[r]{Skewness}}
\put(2057,1697){\makebox(0,0)[r]{SMD}}
\put(2057,1797){\makebox(0,0)[r]{MSD}}
\put(1460,2060){\makebox(0,0){North Pole}}
\put(1460,150){\makebox(0,0){$\alpha$}}
\put(100,1155){%
\special{ps: gsave currentpoint currentpoint translate
270 rotate neg exch neg exch translate}%
\makebox(0,0)[b]{\shortstack{P($\alpha$)/P($\alpha$=0)}}%
\special{ps: currentpoint grestore moveto}%
}
\put(2470,300){\makebox(0,0){1}}
\put(2066,300){\makebox(0,0){0.8}}
\put(1662,300){\makebox(0,0){0.6}}
\put(1258,300){\makebox(0,0){0.4}}
\put(854,300){\makebox(0,0){0.2}}
\put(450,300){\makebox(0,0){0}}
\put(400,1910){\makebox(0,0)[r]{3}}
\put(400,1658){\makebox(0,0)[r]{2.5}}
\put(400,1407){\makebox(0,0)[r]{2}}
\put(400,1155){\makebox(0,0)[r]{1.5}}
\put(400,903){\makebox(0,0)[r]{1}}
\put(400,652){\makebox(0,0)[r]{0.5}}
\put(400,400){\makebox(0,0)[r]{0}}
\end{picture}

%% file: plotsouth.latex
\setlength{\unitlength}{0.1bp}
\special{!
/gnudict 120 dict def
gnudict begin
/Color false def
/Solid false def
/gnulinewidth 5.000 def
/userlinewidth gnulinewidth def
/vshift -33 def
/dl {10 mul} def
/hpt_ 31.5 def
/vpt_ 31.5 def
/hpt hpt_ def
/vpt vpt_ def
/M {moveto} bind def
/L {lineto} bind def
/R {rmoveto} bind def
/V {rlineto} bind def
/vpt2 vpt 2 mul def
/hpt2 hpt 2 mul def
/Lshow { currentpoint stroke M
  0 vshift R show } def
/Rshow { currentpoint stroke M
  dup stringwidth pop neg vshift R show } def
/Cshow { currentpoint stroke M
  dup stringwidth pop -2 div vshift R show } def
/UP { dup vpt_ mul /vpt exch def hpt_ mul /hpt exch def
  /hpt2 hpt 2 mul def /vpt2 vpt 2 mul def } def
/DL { Color {setrgbcolor Solid {pop []} if 0 setdash }
 {pop pop pop Solid {pop []} if 0 setdash} ifelse } def
/BL { stroke gnulinewidth 2 mul setlinewidth } def
/AL { stroke gnulinewidth 2 div setlinewidth } def
/UL { gnulinewidth mul /userlinewidth exch def } def
/PL { stroke userlinewidth setlinewidth } def
/LTb { BL [] 0 0 0 DL } def
/LTa { AL [1 dl 2 dl] 0 setdash 0 0 0 setrgbcolor } def
/LT0 { PL [] 0 1 0 DL } def
/LT1 { PL [4 dl 2 dl] 0 0 1 DL } def
/LT2 { PL [2 dl 3 dl] 1 0 0 DL } def
/LT3 { PL [1 dl 1.5 dl] 1 0 1 DL } def
/LT4 { PL [5 dl 2 dl 1 dl 2 dl] 0 1 1 DL } def
/LT5 { PL [4 dl 3 dl 1 dl 3 dl] 1 1 0 DL } def
/LT6 { PL [2 dl 2 dl 2 dl 4 dl] 0 0 0 DL } def
/LT7 { PL [2 dl 2 dl 2 dl 2 dl 2 dl 4 dl] 1 0.3 0 DL } def
/LT8 { PL [2 dl 2 dl 2 dl 2 dl 2 dl 2 dl 2 dl 4 dl] 0.5 0.5 0.5 DL } def
/Pnt { stroke [] 0 setdash
   gsave 1 setlinecap M 0 0 V stroke grestore } def
/Dia { stroke [] 0 setdash 2 copy vpt add M
  hpt neg vpt neg V hpt vpt neg V
  hpt vpt V hpt neg vpt V closepath stroke
  Pnt } def
/Pls { stroke [] 0 setdash vpt sub M 0 vpt2 V
  currentpoint stroke M
  hpt neg vpt neg R hpt2 0 V stroke
  } def
/Box { stroke [] 0 setdash 2 copy exch hpt sub exch vpt add M
  0 vpt2 neg V hpt2 0 V 0 vpt2 V
  hpt2 neg 0 V closepath stroke
  Pnt } def
/Crs { stroke [] 0 setdash exch hpt sub exch vpt add M
  hpt2 vpt2 neg V currentpoint stroke M
  hpt2 neg 0 R hpt2 vpt2 V stroke } def
/TriU { stroke [] 0 setdash 2 copy vpt 1.12 mul add M
  hpt neg vpt -1.62 mul V
  hpt 2 mul 0 V
  hpt neg vpt 1.62 mul V closepath stroke
  Pnt  } def
/Star { 2 copy Pls Crs } def
/BoxF { stroke [] 0 setdash exch hpt sub exch vpt add M
  0 vpt2 neg V  hpt2 0 V  0 vpt2 V
  hpt2 neg 0 V  closepath fill } def
/TriUF { stroke [] 0 setdash vpt 1.12 mul add M
  hpt neg vpt -1.62 mul V
  hpt 2 mul 0 V
  hpt neg vpt 1.62 mul V closepath fill } def
/TriD { stroke [] 0 setdash 2 copy vpt 1.12 mul sub M
  hpt neg vpt 1.62 mul V
  hpt 2 mul 0 V
  hpt neg vpt -1.62 mul V closepath stroke
  Pnt  } def
/TriDF { stroke [] 0 setdash vpt 1.12 mul sub M
  hpt neg vpt 1.62 mul V
  hpt 2 mul 0 V
  hpt neg vpt -1.62 mul V closepath fill} def
/DiaF { stroke [] 0 setdash vpt add M
  hpt neg vpt neg V hpt vpt neg V
  hpt vpt V hpt neg vpt V closepath fill } def
/Pent { stroke [] 0 setdash 2 copy gsave
  translate 0 hpt M 4 {72 rotate 0 hpt L} repeat
  closepath stroke grestore Pnt } def
/PentF { stroke [] 0 setdash gsave
  translate 0 hpt M 4 {72 rotate 0 hpt L} repeat
  closepath fill grestore } def
/Circle { stroke [] 0 setdash 2 copy
  hpt 0 360 arc stroke Pnt } def
/CircleF { stroke [] 0 setdash hpt 0 360 arc fill } def
/C0 { BL [] 0 setdash 2 copy moveto vpt 90 450  arc } bind def
/C1 { BL [] 0 setdash 2 copy        moveto
       2 copy  vpt 0 90 arc closepath fill
               vpt 0 360 arc closepath } bind def
/C2 { BL [] 0 setdash 2 copy moveto
       2 copy  vpt 90 180 arc closepath fill
               vpt 0 360 arc closepath } bind def
/C3 { BL [] 0 setdash 2 copy moveto
       2 copy  vpt 0 180 arc closepath fill
               vpt 0 360 arc closepath } bind def
/C4 { BL [] 0 setdash 2 copy moveto
       2 copy  vpt 180 270 arc closepath fill
               vpt 0 360 arc closepath } bind def
/C5 { BL [] 0 setdash 2 copy moveto
       2 copy  vpt 0 90 arc
       2 copy moveto
       2 copy  vpt 180 270 arc closepath fill
               vpt 0 360 arc } bind def
/C6 { BL [] 0 setdash 2 copy moveto
      2 copy  vpt 90 270 arc closepath fill
              vpt 0 360 arc closepath } bind def
/C7 { BL [] 0 setdash 2 copy moveto
      2 copy  vpt 0 270 arc closepath fill
              vpt 0 360 arc closepath } bind def
/C8 { BL [] 0 setdash 2 copy moveto
      2 copy vpt 270 360 arc closepath fill
              vpt 0 360 arc closepath } bind def
/C9 { BL [] 0 setdash 2 copy moveto
      2 copy  vpt 270 450 arc closepath fill
              vpt 0 360 arc closepath } bind def
/C10 { BL [] 0 setdash 2 copy 2 copy moveto vpt 270 360 arc closepath fill
       2 copy moveto
       2 copy vpt 90 180 arc closepath fill
               vpt 0 360 arc closepath } bind def
/C11 { BL [] 0 setdash 2 copy moveto
       2 copy  vpt 0 180 arc closepath fill
       2 copy moveto
       2 copy  vpt 270 360 arc closepath fill
               vpt 0 360 arc closepath } bind def
/C12 { BL [] 0 setdash 2 copy moveto
       2 copy  vpt 180 360 arc closepath fill
               vpt 0 360 arc closepath } bind def
/C13 { BL [] 0 setdash  2 copy moveto
       2 copy  vpt 0 90 arc closepath fill
       2 copy moveto
       2 copy  vpt 180 360 arc closepath fill
               vpt 0 360 arc closepath } bind def
/C14 { BL [] 0 setdash 2 copy moveto
       2 copy  vpt 90 360 arc closepath fill
               vpt 0 360 arc } bind def
/C15 { BL [] 0 setdash 2 copy vpt 0 360 arc closepath fill
               vpt 0 360 arc closepath } bind def
/Rec   { newpath 4 2 roll moveto 1 index 0 rlineto 0 exch rlineto
       neg 0 rlineto closepath } bind def
/Square { dup Rec } bind def
/Bsquare { vpt sub exch vpt sub exch vpt2 Square } bind def
/S0 { BL [] 0 setdash 2 copy moveto 0 vpt rlineto BL Bsquare } bind def
/S1 { BL [] 0 setdash 2 copy vpt Square fill Bsquare } bind def
/S2 { BL [] 0 setdash 2 copy exch vpt sub exch vpt Square fill Bsquare } bind def
/S3 { BL [] 0 setdash 2 copy exch vpt sub exch vpt2 vpt Rec fill Bsquare } bind def
/S4 { BL [] 0 setdash 2 copy exch vpt sub exch vpt sub vpt Square fill Bsquare } bind def
/S5 { BL [] 0 setdash 2 copy 2 copy vpt Square fill
       exch vpt sub exch vpt sub vpt Square fill Bsquare } bind def
/S6 { BL [] 0 setdash 2 copy exch vpt sub exch vpt sub vpt vpt2 Rec fill Bsquare } bind def
/S7 { BL [] 0 setdash 2 copy exch vpt sub exch vpt sub vpt vpt2 Rec fill
       2 copy vpt Square fill
       Bsquare } bind def
/S8 { BL [] 0 setdash 2 copy vpt sub vpt Square fill Bsquare } bind def
/S9 { BL [] 0 setdash 2 copy vpt sub vpt vpt2 Rec fill Bsquare } bind def
/S10 { BL [] 0 setdash 2 copy vpt sub vpt Square fill 2 copy exch vpt sub exch vpt Square fill
       Bsquare } bind def
/S11 { BL [] 0 setdash 2 copy vpt sub vpt Square fill 2 copy exch vpt sub exch vpt2 vpt Rec fill
       Bsquare } bind def
/S12 { BL [] 0 setdash 2 copy exch vpt sub exch vpt sub vpt2 vpt Rec fill Bsquare } bind def
/S13 { BL [] 0 setdash 2 copy exch vpt sub exch vpt sub vpt2 vpt Rec fill
       2 copy vpt Square fill Bsquare } bind def
/S14 { BL [] 0 setdash 2 copy exch vpt sub exch vpt sub vpt2 vpt Rec fill
       2 copy exch vpt sub exch vpt Square fill Bsquare } bind def
/S15 { BL [] 0 setdash 2 copy Bsquare fill Bsquare } bind def
/D0 { gsave translate 45 rotate 0 0 S0 stroke grestore } bind def
/D1 { gsave translate 45 rotate 0 0 S1 stroke grestore } bind def
/D2 { gsave translate 45 rotate 0 0 S2 stroke grestore } bind def
/D3 { gsave translate 45 rotate 0 0 S3 stroke grestore } bind def
/D4 { gsave translate 45 rotate 0 0 S4 stroke grestore } bind def
/D5 { gsave translate 45 rotate 0 0 S5 stroke grestore } bind def
/D6 { gsave translate 45 rotate 0 0 S6 stroke grestore } bind def
/D7 { gsave translate 45 rotate 0 0 S7 stroke grestore } bind def
/D8 { gsave translate 45 rotate 0 0 S8 stroke grestore } bind def
/D9 { gsave translate 45 rotate 0 0 S9 stroke grestore } bind def
/D10 { gsave translate 45 rotate 0 0 S10 stroke grestore } bind def
/D11 { gsave translate 45 rotate 0 0 S11 stroke grestore } bind def
/D12 { gsave translate 45 rotate 0 0 S12 stroke grestore } bind def
/D13 { gsave translate 45 rotate 0 0 S13 stroke grestore } bind def
/D14 { gsave translate 45 rotate 0 0 S14 stroke grestore } bind def
/D15 { gsave translate 45 rotate 0 0 S15 stroke grestore } bind def
/DiaE { stroke [] 0 setdash vpt add M
  hpt neg vpt neg V hpt vpt neg V
  hpt vpt V hpt neg vpt V closepath stroke } def
/BoxE { stroke [] 0 setdash exch hpt sub exch vpt add M
  0 vpt2 neg V hpt2 0 V 0 vpt2 V
  hpt2 neg 0 V closepath stroke } def
/TriUE { stroke [] 0 setdash vpt 1.12 mul add M
  hpt neg vpt -1.62 mul V
  hpt 2 mul 0 V
  hpt neg vpt 1.62 mul V closepath stroke } def
/TriDE { stroke [] 0 setdash vpt 1.12 mul sub M
  hpt neg vpt 1.62 mul V
  hpt 2 mul 0 V
  hpt neg vpt -1.62 mul V closepath stroke } def
/PentE { stroke [] 0 setdash gsave
  translate 0 hpt M 4 {72 rotate 0 hpt L} repeat
  closepath stroke grestore } def
/CircE { stroke [] 0 setdash 
  hpt 0 360 arc stroke } def
/BoxFill { gsave Rec 1 setgray fill grestore } def
end
}
\begin{picture}(2520,2160)(0,0)
\special{"
gnudict begin
gsave
0 0 translate
0.100 0.100 scale
0 setgray
newpath
LTb
450 400 M
63 0 V
1957 0 R
-63 0 V
450 652 M
63 0 V
1957 0 R
-63 0 V
450 903 M
63 0 V
1957 0 R
-63 0 V
450 1155 M
63 0 V
1957 0 R
-63 0 V
450 1407 M
63 0 V
1957 0 R
-63 0 V
450 1658 M
63 0 V
1957 0 R
-63 0 V
450 1910 M
63 0 V
1957 0 R
-63 0 V
450 400 M
0 63 V
0 1447 R
0 -63 V
854 400 M
0 63 V
0 1447 R
0 -63 V
1258 400 M
0 63 V
0 1447 R
0 -63 V
1662 400 M
0 63 V
0 1447 R
0 -63 V
2066 400 M
0 63 V
0 1447 R
0 -63 V
2470 400 M
0 63 V
0 1447 R
0 -63 V
LTb
450 400 M
2020 0 V
0 1510 V
-2020 0 V
450 400 L
1.000 UL
LT0
2107 1797 M
263 0 V
450 903 M
16 2 V
17 2 V
16 3 V
16 5 V
17 7 V
16 10 V
16 12 V
17 17 V
16 21 V
16 26 V
17 32 V
16 37 V
16 44 V
17 49 V
16 54 V
16 56 V
16 56 V
17 56 V
16 52 V
16 49 V
17 43 V
16 34 V
16 26 V
17 15 V
16 2 V
16 -12 V
17 -23 V
16 -34 V
16 -43 V
17 -51 V
16 -56 V
16 -62 V
17 -64 V
16 -66 V
16 -67 V
17 -65 V
16 -62 V
16 -58 V
17 -54 V
16 -50 V
16 -45 V
17 -42 V
16 -38 V
16 -35 V
17 -31 V
16 -28 V
16 -25 V
17 -22 V
16 -19 V
16 -17 V
16 -14 V
17 -12 V
16 -10 V
16 -8 V
17 -7 V
16 -6 V
16 -5 V
17 -4 V
16 -3 V
16 -4 V
17 -3 V
16 -3 V
16 -4 V
17 -4 V
16 -5 V
16 -4 V
17 -5 V
16 -5 V
16 -5 V
17 -5 V
16 -4 V
16 -5 V
17 -4 V
16 -4 V
16 -3 V
17 -3 V
16 -3 V
14 -2 V
248 0 R
15 1 V
17 2 V
16 2 V
16 1 V
17 2 V
16 2 V
1.000 UL
LT1
2107 1697 M
263 0 V
450 903 M
16 7 V
17 7 V
16 7 V
16 8 V
17 9 V
16 11 V
16 12 V
17 14 V
16 16 V
16 18 V
17 21 V
16 24 V
16 27 V
17 29 V
16 32 V
16 32 V
16 32 V
17 31 V
16 30 V
16 27 V
17 23 V
16 19 V
16 14 V
17 7 V
16 0 V
16 -7 V
17 -14 V
16 -20 V
16 -26 V
17 -31 V
16 -34 V
16 -38 V
17 -40 V
16 -42 V
16 -44 V
17 -43 V
16 -44 V
16 -42 V
17 -41 V
16 -40 V
16 -38 V
17 -37 V
16 -34 V
16 -33 V
17 -31 V
16 -28 V
16 -26 V
17 -23 V
16 -21 V
16 -18 V
16 -16 V
17 -12 V
16 -11 V
16 -8 V
17 -7 V
16 -6 V
16 -4 V
17 -3 V
16 -3 V
16 -2 V
17 -2 V
16 -2 V
16 -3 V
17 -2 V
16 -3 V
16 -4 V
17 -3 V
16 -4 V
16 -3 V
17 -4 V
16 -3 V
16 -4 V
17 -3 V
16 -4 V
16 -3 V
17 -3 V
16 -3 V
16 -2 V
17 -3 V
16 -2 V
16 -2 V
17 -2 V
16 -2 V
16 -2 V
16 -1 V
17 -2 V
16 -1 V
16 -2 V
17 -1 V
16 -1 V
16 -1 V
17 -1 V
16 -1 V
16 -1 V
17 -1 V
16 -1 V
16 0 V
17 -1 V
16 -1 V
1.000 UL
LT2
2107 1597 M
263 0 V
450 903 M
31 -21 V
30 -20 V
31 -17 V
30 -13 V
31 -9 V
31 -1 V
30 6 V
31 14 V
30 19 V
31 22 V
31 21 V
30 20 V
31 14 V
30 6 V
31 0 V
31 -5 V
30 -8 V
31 -10 V
31 -10 V
30 -7 V
31 -5 V
30 -2 V
31 2 V
31 3 V
30 7 V
31 8 V
30 11 V
31 12 V
31 12 V
30 12 V
31 9 V
30 6 V
31 2 V
31 -3 V
30 -8 V
31 -9 V
30 -10 V
31 -8 V
31 -6 V
30 -2 V
31 3 V
30 8 V
31 11 V
31 14 V
30 16 V
31 19 V
30 19 V
31 19 V
31 18 V
30 18 V
31 15 V
31 14 V
30 9 V
31 6 V
30 3 V
31 0 V
31 -3 V
30 -5 V
31 -7 V
30 -8 V
31 -10 V
31 -11 V
30 -11 V
31 -11 V
30 -10 V
31 -10 V
1.000 UL
LT3
2107 1497 M
263 0 V
450 903 M
31 -2 V
30 -2 V
31 -3 V
30 -5 V
31 -5 V
31 -7 V
30 -9 V
31 -10 V
30 -12 V
31 -10 V
31 -9 V
30 -6 V
31 -2 V
30 3 V
31 6 V
31 9 V
30 9 V
31 7 V
31 5 V
30 -1 V
31 -7 V
30 -11 V
31 -14 V
31 -15 V
30 -14 V
31 -12 V
30 -9 V
31 -5 V
31 -3 V
30 -2 V
31 -3 V
30 -5 V
31 -8 V
31 -12 V
30 -16 V
31 -17 V
30 -19 V
31 -19 V
31 -17 V
30 -14 V
31 -11 V
30 -9 V
31 -5 V
31 -4 V
30 -2 V
31 0 V
30 0 V
31 1 V
31 2 V
30 1 V
31 1 V
31 -1 V
30 -1 V
31 -3 V
30 -5 V
31 -5 V
31 -6 V
30 -7 V
31 -7 V
30 -7 V
31 -7 V
31 -7 V
30 -5 V
31 -5 V
30 -4 V
31 -2 V
stroke
grestore
end
showpage
}
\put(2057,1497){\makebox(0,0)[r]{Kurtosis}}
\put(2057,1597){\makebox(0,0)[r]{Skewness}}
\put(2057,1697){\makebox(0,0)[r]{SMD}}
\put(2057,1797){\makebox(0,0)[r]{MSD}}
\put(1460,2060){\makebox(0,0){South Pole}}
\put(1460,150){\makebox(0,0){$\alpha$}}
\put(100,1155){%
\special{ps: gsave currentpoint currentpoint translate
270 rotate neg exch neg exch translate}%
\makebox(0,0)[b]{\shortstack{P($\alpha$)/P($\alpha$=0)}}%
\special{ps: currentpoint grestore moveto}%
}
\put(2470,300){\makebox(0,0){1}}
\put(2066,300){\makebox(0,0){0.8}}
\put(1662,300){\makebox(0,0){0.6}}
\put(1258,300){\makebox(0,0){0.4}}
\put(854,300){\makebox(0,0){0.2}}
\put(450,300){\makebox(0,0){0}}
\put(400,1910){\makebox(0,0)[r]{3}}
\put(400,1658){\makebox(0,0)[r]{2.5}}
\put(400,1407){\makebox(0,0)[r]{2}}
\put(400,1155){\makebox(0,0)[r]{1.5}}
\put(400,903){\makebox(0,0)[r]{1}}
\put(400,652){\makebox(0,0)[r]{0.5}}
\put(400,400){\makebox(0,0)[r]{0}}
\end{picture}